# DEADLOCK RECOVERY TECHNIQUE IN BUS ENHANCED NOC ARCHITECTURE

Saeid Sharifian Nia[1], Abbas Vafaei[2], Hamid Shahimohamadi[3]

[1]Department of Computer Engineering, University of Isfahan, Isfahan, Iran
s.sharifian@eng.ui.ac.ir

[2]Department of Computer Engineering, University of Isfahan, Isfahan, Iran
abbas_vafaei@eng.ui.ac.ir

[3]Department of Computer Engineering, Shahid Beheshti University, Tehran, Iran
h.shahi@mail.sbu.ac.ir

## ABSTRACT

*Increase in the speed of processors has led to crucial role of communication in the performance of systems. As a result, routing is taken into consideration as one of the most important subjects of the Network on Chip architecture. Routing algorithms to deadlock avoidance prevent packets route completely based on network traffic condition by means of restricting the route of packets. This action leads to less performance especially in non-uniform traffic patterns. On the other hand True Fully Adoptive Routing algorithm provides routing of packets completely based on traffic condition. However, deadlock detection and recovery mechanisms are needed to handle deadlocks. Use of global bus beside NoC as a parallel supportive environment, provide platform to offer advantages of both features of bus and NoC. This bus is useful for broadcast and multicast operations, sending delay sensitive signals, system management and other services. In this research, we use this bus as an escaping path for deadlock recovery technique. According to simulation results, this bus is suitable platform for deadlock recovery technique.*

## KEYWORDS

*Network on chip, deadlock recovery, deadlock detection, routing algorithm, bus enhanced NoC.*

## 1. INTRODUCTION

Among most important predictions of IC manufacturing is Moore rule. This rule indicates the fast growth of integrated elements in the ICs. Use of the bus beside NoC as a synergetic media, improve existing functionality and offering new services. One of the most important applicable of this bus is broadcast and multicast operations. This bus provides these operations without using broadcast and multicast protocols on the network. We use this synergetic media as an escaping path for deadlock recovery technique in the network.

Deadlock recovery and avoidance are two most important strategies for deadlock handling in interconnection networks. A deadlock occurs when some packets cannot advance to their destination in a cyclic form because all packets in the cycle hold their resources and request other resources that are occupied by another packet. Deadlock cannot occur if there is not any cyclic dependency between channels. Deadlock avoidance techniques restrict routing algorithms in order to prevent this cycle. These restrictions degrade performance especially in non-uniform traffic patterns. Deadlock recovery techniques provide flexible routing. Packets route in network only based on traffic condition without any restriction. This routing algorithm is called True Fully Adaptive Routing (TFAR). Deadlock recovery techniques are different in detection and recovery mechanisms.

In [2] is shown that deadlock rarely occurs when sufficient routing is provided (e.g. TFAR) but it occurs more when the network is closed or beyond saturation. Use of injection limitation mechanisms can prevent this situation [7, 8]. Another way to reduce deadlock occuration is use of virtual channels.

According to how a path is defined, routing algorithms are classified as deterministic or adaptive. In deterministic algorithms the path between the source and destination are always the same. Adaptive routing algorithms may considerably improve performance over deterministic routing algorithms, especially in non-uniform traffic. However, adaptive routers are more complex, slowing down clock frequency. This algorithms use information about network traffic to avoid congestion or faulty regions of network [1].

Adaptive routing algorithms in order to avoid deadlock use some restrictions on routing. According to these restrictions, adaptive routing algorithms are classified to fully adaptive or partially adaptive. In partially adaptive routing, in order to avoid deadlock, two turns of routes are forbidden. The well-known of this algorithms are Minimal West First, Minimal North Last, Odd-Even, etc. for example in Minimal West First, if destination node is in west of source, flits must route in west direction to arrive x coordinate of destination node then route to north or south toward destination node. If destination node is not in west of source, flits can route adaptively in minimal path to east, north or south directions [3-5].

In fully adaptive routing algorithms in order to avoid deadlock, routing algorithms restrict use of virtual channels in each direction [1]. These techniques need a number of virtual channels in each router direction and may not beneficially use resources. TFAR algorithm provides routing without restriction on the use of virtual channels and turns in network but it needs to use detection and recovery mechanism in network.

In this paper we will introduce a new technique of deadlock recovery. In section 2, presents some deadlock detection and recovery mechanisms. Section 3 is briefly present advantage of use bus as an internal part of NoC. We discuss our mechanism in section 4. Section 5 gives the performance results of true fully adaptive routing with our proposed deadlock recovery technique. Finally section 6 presents some conclusions and future works.

## 2. RELATED WORKS

The deadlock detection mechanism in [9] is based on measuring the time when the header flit is blocked. Probability of false detections in this mechanism is high. Because a blocked message maybe is waiting for channels that are occupied with passing long messages. In this situation the header flit is blocked for a nmber of clock cycles, but the output channels that header requested are not blocked because they are passing flits from other input channels of router. Clearly, this mechanism dose not show whether output channel is active or not.

 A more efficient mechanism measures the time that inactive channels which have been occupied with blocked message are requested by a message. To recognize a blocked message, the router needs to monitor activity of each channel. The monitoring done by checking, one bit of a counter as a flag. A message is presumed to be deadlock only if all output virtual channels that are requested by that message contain blocked messages for a period of time. When a routing algorithm can choose all virtual channels in each physical channel to route a message (e.g. TFAR), it is only necessary to monitor the activity of physical channels instead of checking the activity of virtual channels.

For deadlock detection mechanism, in the routers every output physical channel is associated with a counter. This counter is incremented every clock cycle and resets when a flit is transmitted across the physical channel. The counter contains cycles that all output virtual channels in that direction are inactive. If all requested output channels are busy, the value of timers is checked and if it is greater than predetermine value of threshold, router presumes that the message is in deadlock.

Performance depends directly on choosing the value of threshold. If it chooses the great value, the realized packets in deadlock must wait for this threshold time to detect and if the small value is chosen many false deadlocks are detected. To reduce false detection in [11], they used local traffic information to choose the value of threshold dynamically.

Deadlock recovery techniques are classified in regressive and progressive techniques. Regressive techniques kill deadlock flits in intermediate routers and re inject the deadlock message at original source node. In proposed regressive technique [12], the detection mechanism is done in source node. When a header flit injects to network, a counter starts and by injecting each flit in network counter is restarted. If the counter reaches threshold value before the last flit has been injected to the network it is presumed that the deadlock occurs and flits are removed from routers and the packet is re injected again. Removing the flits is done by sending a control signal from source node along the path that is reserved by the header flit.

If the tail flit injects, it is guaranteed that header flit have been arrived to the destination without deadlock occurring. If length of packet is smaller than the number of pads it is necessary that source node adds some null flits at the end of the packet until the header reaches to destination. If many short packets are being sent, maximum throughput can be significantly reduced. If deadlock detection mechanism is done in source node, regressive deadlock recovery technique is usually used [1].

More efficient deadlock recovery techniques are progressive recovery. Deadlock recovery techniques according to needs of recourses are classified to hardware and software. Well known hardware based progressive deadlock recovery is Disha [9]. Disha uses central buffer in each router that is called Deadlock Buffer. This buffer is not allocated to any special direction in the router and all input channels can use it. All these deadlock buffers in network can be viewed as virtual network which can be used to route the deadlock messages. Accessing to this virtual network must be mutual exclusive and it is implemented with a circular token. When a router detects a deadlock message and the router has the token, router directs this message to Deadlock Buffer. This message routes to Deadlock Buffers in network until reaches to destination. At each router flits in Deadlock Buffer have a higher priority than other output channels for routing.

Other progressive deadlock recovery technique is software based. In software based deadlock recovery [10] after detection mechanism, router ejects deadlock message - whose header flit is in head of input buffer - to current node. And re inject this message into the network at a later time. This strategy needs a software messaging layer to recognize foreign messages. Also it needs some buffer space in the local nodes to eject a deadlock message completely from network in order to open the input channel for passing another message in deadlock cycle. A round robin policy between ejected message and locally generated message is used to select witch message must be injected into the network.

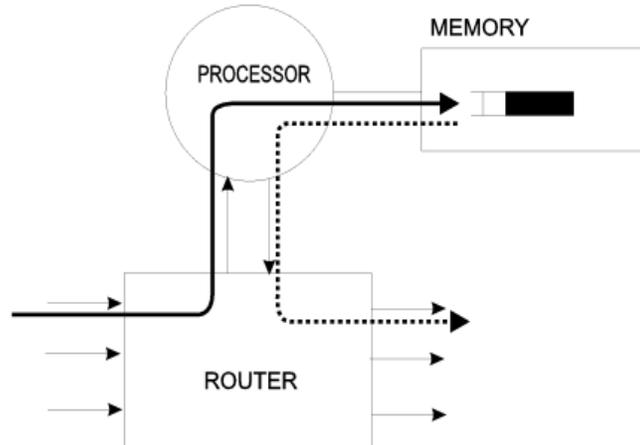

Figure 1: Software based deadlock recovery mechanism [10]

In [10] message injection limitation is introduced as a mechanism to reduce deadlock probability and performance degridation. These mechanisms prevent that network from tending to saturation. They use the number of busy channels in the router as a paragon to prevent network from tending to saturation. When the number of busy output channels exceed a threshold value the router prevents injection of new message. By choosing apropriate value of threshold the network is immune to tend toward saturation. This technique is more applicable when network uses several virtual channels in each direction of routers.

In a deadlock cycle at least the header flit of one message is at head of an input buffer of the router [10]. This header flit requests routing but cannot route due to deadlock. It is desirable that deadlock recovery technique ejects one of the messages in deadlock cycle.

## 3. ARCHITECTURE

Use of a global bus inside NoC provides platform to offer advantages of both features of bus and NoC. The most basic service provided by the bus is broadcast operations. With implemented of segmented bus, scalability and multicast operations is provided [13]. Also the bus inside NoC can be used for latency sensitive signalling instead of sending signals in a multi hop switching platform. In [13-17] they describe some of the advantage of bus and use of it inside NoC. We use this bus as an escaping path for recovery mechanism in network.

## 4. PROPOSED MECHANISM

True Fully Adaptive Routing algorithm, use routes between source and destination nodes more beneficially especially in non-uniform traffic patterns. We use deadlock detection mechanism [10], as discussed in part 2, for each output physical channel of routers a counter is used. If a flit is transmitted across a physical channel the counter resets and it is incremented every clock cycle. Therefore the counters indicate the number of cycles that physical channel is inactive. According to figure 2, checking one bit of counter as a flag is sufficient to find weather the timer reaches to the threshold value or not.

When a header flit cannot route -because all directions that header flit requested for routing are reserved by other messages in input channels of the router- and all values of counters that header requested exceed predetermined value, router presumes that message is in deadlock and recovery phase is started.

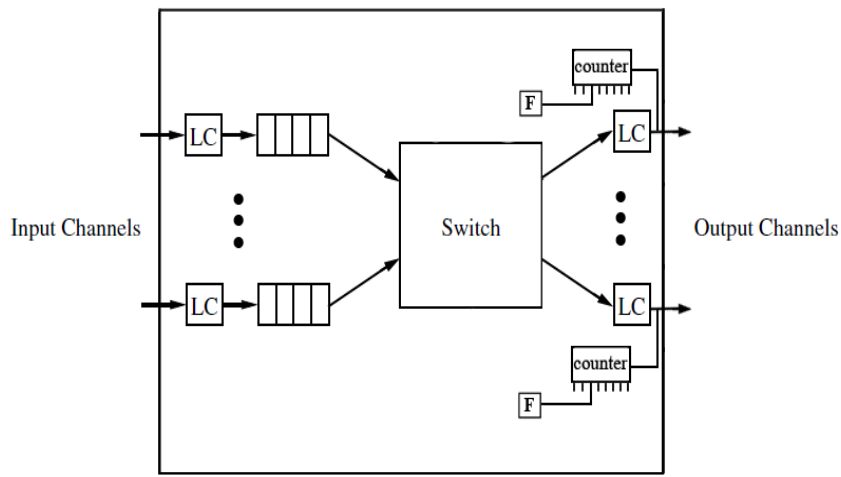

Figure 2: deadlock detection mechanism

Our hardware deadlock recovery mechanism uses a global bus to escape messages from deadlock. After detection mechanism for recovery, router sends a request to bus arbiter. The bus arbiter collects all Bus Request (BR) signals in a queue. If the bus is free, arbiter sends the Bus Grant (BG) signal to router that its request is in the head of the queue.

When a router receives bus grant, forwards the header flit in Input Buffer (IB) of bus. All Processing Elements (PE) can monitor this header flit at Output Buffers (OB) of bus. If destination address of header flit is the same as address of processing element, the processing element picks up this header and waits for receiving next flit. Until the tail flit arrives, the processing element continues picking up flits. Thoroughly the other processing elements monitor output buffer of bus to receive header flit of one packet.

Each router cancels its bus request when forwarding the tail flit of deadlock message on the bus. It is possible that a header flit that is waiting for receiving bus grant from bus arbiter be routed in the router. It is because the output channel that header flit requests has gotten free. In this situation the message routes in the router and router sends cancellation signal of bus request to the bus arbiter.

When a bus arbiter receives cancellation signal from a router, removes the number of router from the queue. If the bus arbiter removes the number of router from head of the queue, it sends a bus grant to the router whose request is coming in head of the queue.

When a processing element receives the header flit from output buffer of bus with the same destination address as its own, it is probable that processing element is receiving flits of a message from the router. In this situation the header flit at output buffer of bus must wait until the processing element receives the tail flit from the router. After that the processing element starts to get flits from bus.

However if requests from bus and router arrive simultaneously to the processing element, the header flit witch receives from bus is prior than the other one. In this situation the processing element picks up the flits from bus until the tail flit receives and then starts to get flits from the router. By arrival of the tail flit from bus to the processing element, the router sends a cancellation signal of bus request to the bus arbiter. The priority of receiving the requests from bus is for the bus to get free in a faster rate and to give services to other routers that request using of bus. If deadlock recovery technique cannot escape deadlock message fast enough, the average of latency of packets in network will be increased. Also the blocked message occupies the bandwidth and therefor decreases throughput of network

According to figure 3, the proposed router compared to input-buffered router with no virtual channels [1] has capabilities such as sending data on the bus, sending request signal to bus arbiter and receiving grant signal from arbiter. Link Controller (LC) performs flow of messages across the physical channel between adjacent routers. It is needed to put link controller on either side of a channel to coordinate transfer units of flow control [1].

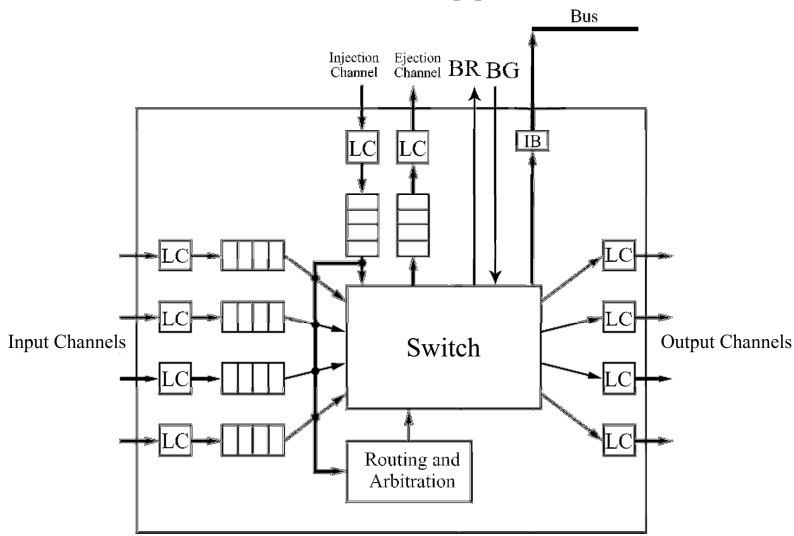

Figure 3: Proposed router architecture

In our proposed architecture the size of input and output buffers of bus is only one flit. The router forward flits to input buffer of bus and on the other hand PEs can monitor flits on the output buffers. The size of input and output buffers of bus leads to decreasing the power consumption as compared to common bus.

There are several packets in every cycle of deadlock. In deadlock detection and recovery mechanisms, it is desirable to escape only one of the messages from deadlock cycle. Escaping one packet from deadlock, allow other messages to route through the routers. In our proposed mechanism, only one message can use the bus in a time. So by forwarding flits of one message in deadlock cycle on the bus, other messages in deadlock cycle can route. Therefore they deny their requests for using the bus. Accordingly, deadlock recovery is done only for one message in the deadlock cycle.

## 5. SIMULATION RESULTS

The work consists of simulation of bus, detection mechanism and recovery by means of forwarding flits of deadlocked message on the bus. For evaluation of the results, we use Noxim simulator. These results are based on 2-Dimension 4x4 mesh topology. The packets length is between 4 to 10 flits. It is simulated under non-uniform traffic loads including First Matrix Transpose, Butterfly and Bit Reversal.

We compare our deadlock recovery technique - true fully adaptive recovery with bus - with three different routing algorithms XY, Minimal West First and Odd-Even. As shown in figure 4, we compared average of packets latency and throughput metrics with increasing packet injection rate. Our proposed deadlock recovery mechanism with increase packets injection rate keep the value of latency of packets in lower than other routing algorithms in First Matrix Transpose traffic pattern. And with better use of routes between sources and destinations the average of throughput is increased, compared to other routing algorithms.

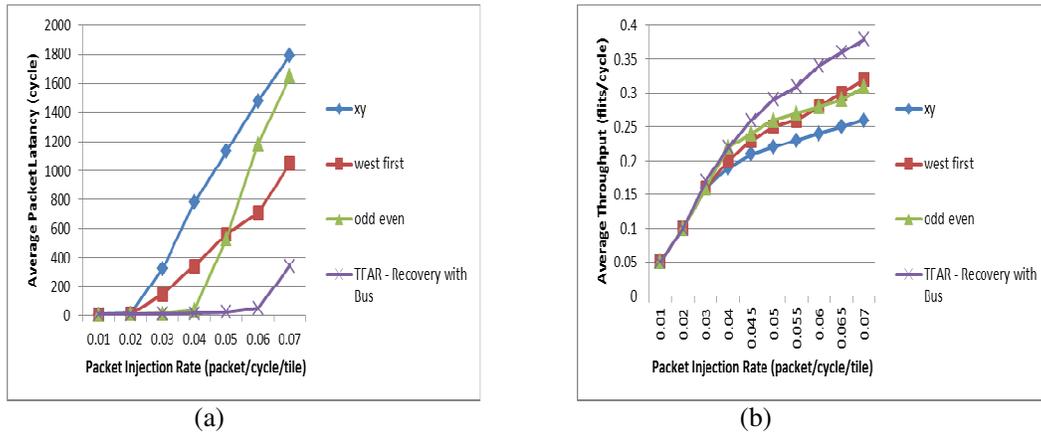

(a)                          (b)

Figure 4: Average of Packat Latency and Throughput in First Matrix Transpose traffic pattern.

However in figure 5, we compared the latency of packets in Butterfly and Bit Reversal traffic patterns. According to figure 5 with lower amounts of average latency, our proposed mechanism has provided more increase, packet injection rate, as compared to other routing algorithms.

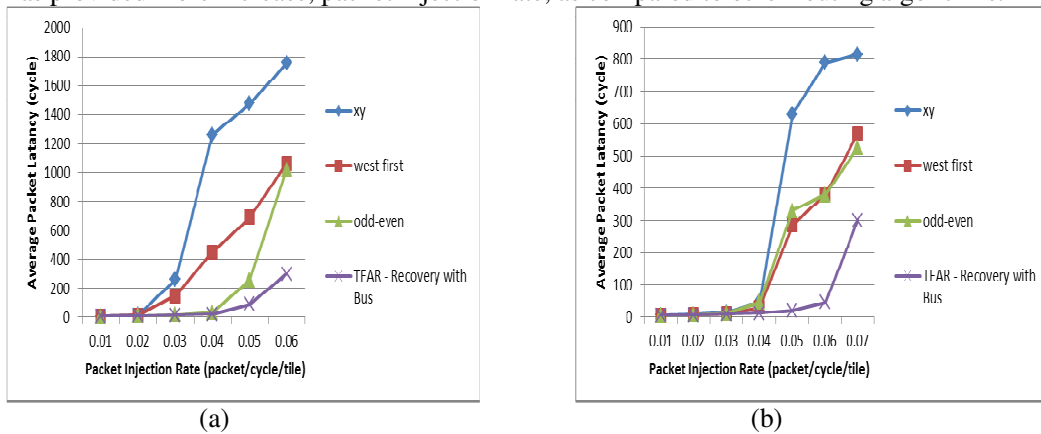

(a)                          (b)

Figure 5: (a) Average of Packat Latency in Bit Reversal traffic pattern. (b) Average of Packat Latency in Butterfly traffic pattern.

## 6. CONCLUSIONS

Increase in the speed of processors has led to important role of communications in interconnection networks. The restrictions that deadlock avoidance routing algorithms apply on the routing of packets prevent the packet to be routed completely base on network traffic condition. The True Fully Adaptive Routing algorithm provides packets routing completely base on traffic condition. A bus adjacent NoC improve the performance of network and provides the bus advantage beside NoC. The simulation results are shown, this bus suitable for deadlock recovery.

According to deadlock rarely occurrence, when the network is not close or beyond saturation if flexible routing algorithm is used, this bus is applicable for broadcast and multicast operations, system management, delay sensitive signals and etc.

With increase of packet injection rate, the network tends to saturation. Therefore latency of packets in reaching to destination will severely increase with respect to algorithms adaptation and traffic patterns. Adding virtual channels in each direction in routers can increase network throughput. Also uses of two virtual channels per physical channel have been shown to be enough to reduce probability of deadlock to very small values. Our future objections are discussion of the effect of virtual channels on average of throughput and packets latency in the architecture of network on chip with enhanced bus.